# Enhanced spectral range of strain-induced tuning of quantum dots in circular Bragg grating cavities

*Ivan Gamov,\* Matthias Sauter, Samuel Huber, Quirin Buchinger, Peter Gschwandtner, Ulrike Wallrabe, Sven Höfling, Tobias Huber-Loyola*

\*e-mail: ivan.gamov@imtek.uni-freiburg.de

*Ivan Gamov, Matthias Sauter, Samuel Huber, Quirin Buchinger, Peter Gschwandtner, Sven Höfling, Tobias Huber-Loyola*

Julius-Maximilians-Universität Würzburg, Physikalisches Institut and Würzburg-Dresden Cluster of Excellence ct.qmat, Lehrstuhl für Technische Physik, Am Hubland, 97074 Würzburg, Germany

*Ivan Gamov, Ulrike Wallrabe*

Laboratory for Microactuators, Department of Microsystems Engineering – IMTEK, University of Freiburg, Georges-Koehler-Allee 102, 79110 Freiburg, Germany

*Samuel Huber*

Centre for Nanosciences and Nanotechnologies CNRS, Universite Paris-Saclay, UMR 9001, 91120 Palaiseau, France

*Tobias Huber-Loyola*

Institute of Photonics and Quantum Electronics (IPQ), Karlsruhe Institute of Technology (KIT), Karlsruhe, Germany

Funding: Bavaria and the Federal Ministry of Research, Technology and Space (BMFTR) within Projects QR.X (FKZ: 16KISQ010), Qecs (FKZ: 13N16272) and QR.N (FKZ: 16KIS2209); Vector Stiftung (Grant No. P2024-0772), Baden-Württemberg, Germany.

**Abstract**

Tunable sources of entangled and single photons are essential for implementing entanglement-based quantum information protocols, as quantum teleportation and entanglement swapping depend on photon indistinguishability. Tunable devices are fabricated from indium arsenide (InAs) quantum dots (QDs) embedded in gallium arsenide (GaAs) nanomembranes placed on monolithic piezoelectric substrates. Circular Bragg grating (CBG) resonators enhance emission brightness and exploit the Purcell effect; however, the inclusion of CBGs reduces strain-mediated tunability compared to planar nanomembranes. A simple and effective solution is introduced: filling the CBG trenches with a stiff dielectric (aluminum oxide, $Al_2O_3$) via atomic layer deposition (ALD) restores up to 95% of the tunability of planar structures. Finite element analysis (FEA) confirms that the tunability loss originates from bending in the device layers due to strain relief in the CBG geometry. Lowering the stiffness of intermediate layers between the QDs and the piezoelectric actuator, such as in bonding or reflector layers, further increases strain losses in uncoated CBGs. Coated devices maintain 98–99% strain-tuning efficiency across all simulated underlayer stiffnesses. The results demonstrate that advantageous optical cavity properties can be effectively combined with piezoelectric strain tuning, enabling scalable, bright, and tunable quantum light sources.

# 1. Introduction

Photons are ideal carriers of quantum information due to their weak interaction with the environment and their compatibility with sending them through air or glass fibers with low losses.[1,2] Among the numerous material platforms explored for creating quantum light sources,[3–7] semiconductor quantum dots (QDs) stand out as particularly promising.[8–10] However, creating a bright and tunable photon source has proven to be more challenging in practice than anticipated, due to the need to simultaneously achieve high photon flux and high indistinguishability. These challenges can be effectively addressed through the integration of QDs into microcavities of different geometries. Among various cavity designs,[11,12] circular Bragg grating resonators (CBGs) provide a high Purcell factor over a broad emission range,[13,14] enhancing emission rates and enabling efficient collection of multiple QD transitions. [15,16] The emission wavelength of QDs can be adjusted through growth conditions[17] or by choosing materials[18–20] but tuning on the device level is necessary to overcome growth fluctuations for quantum light applications. Several methods have been developed to enable spectral tunability, such as temperature tuning,[7,21,22] optical tuning,[23,24] electric field,[25–27] or magnetic field tuning.[25,28,29] Another approach involves strain-based spectral tuning. It allows for an extended tuning range while maintaining high photon coherence and high photon extraction probability[30–35]. Monolithic piezoelectric substrates without structuring provide both high robustness and high tunability. For thinned interfaces between the actuator and QDs, spectral shift over 10 meV (12 nm) was demonstrated.[30] Fine structure splitting (FSS) strain tuning required for multiphoton operation was shown for micromachined piezo substrates and can also be achieved with planar actuators.[36] However, strain transfer can struggle in cavity systems when the latter is not optimized for strain tuning. For example, it was shown that strain tuning in micropillar cavities is ineffective since the actuator does not transfer strain efficiently to QDs but instead the strain relaxes in the freestanding structure due to their high aspect ratio.[33] When integrating CBGs onto piezoelectric substrates, tunability has been shown;[32,33] however, the strain transfer waned in the cavity configuration, unlike in its non-cavity counterpart.

In this paper, we demonstrate the tunability enhancement of QDs embedded into CBGs on monolithic piezoelectric substrates. Simulations of conventional CBG structures reveal a drop of tunability due to strain relief, when the etched trenches of the CBG disrupt the continuity of the monolithic membrane. On the way to a QD, over 50% of the generated strain can be lost. An additional layer of aluminum oxide ($Al_2O_3$) deposited on top of CBGs is shown to be an effective treatment for restoring the original high tunability of the sample in both experiment and simulation. According to measurements, the strain transfer during the actuation rises from 31±4% in the standard CBGs with air trenches to 91±4% when depositing $Al_2O_3$, relative to the tunability in the monolithic membranes (100%). Simulations of strain in both the configurations (air vs. $Al_2O_3$) help to understand the principles of the strain distribution in layers and help to find the most impactful parameters of materials in the device fabrication.

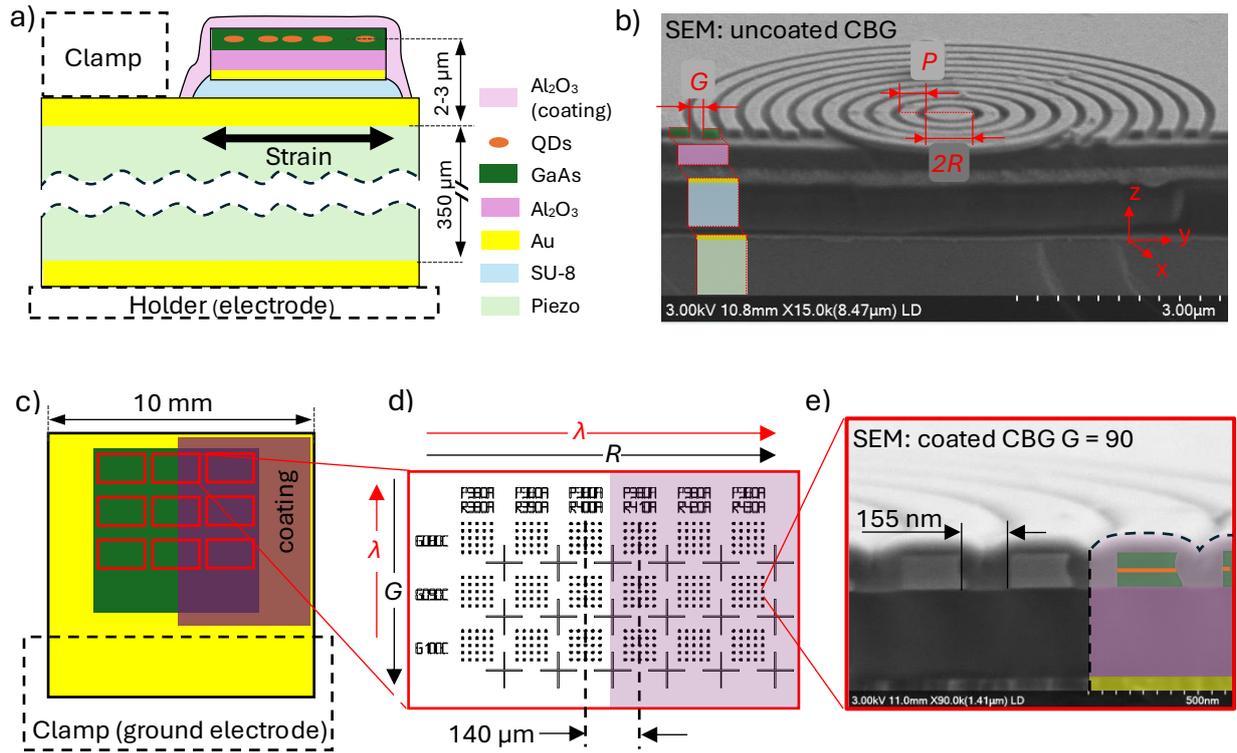

**Figure 1.** (a) Schematic of the sample layers in the cross-section indicated by colors and (b) SEM image of a irregular fracture at uncoated CBG with coordinate system and definition of $R, P, G$ processing parameters. The colored sections on the left show the local cleave in z-y plane with assumed offset along x-axis. (c) On the top view, the right half of the sample is coated with $Al_2O_3$. The CBG layout includes 9 fields (red rectangles) with the same variation of the parameters. d) Each single field includes 450 CBGs at parameter variation of $R$ and $G$ to control the resonance wavelength $\lambda$ of the CBGs. The distance between neighboring CBGs within the same parameters (squares 5 by 5 devices) is 20 µm and the distance between the closest coated to uncoated CBGs is 140 µm. The SEM image (e) shows the cross-section of a coated CBG at $G$ = 90 nm.

## 2. Measurements of photoluminescence spectral shift

The spectral shift of quantum dot photoluminescence is measured for III–V QD devices integrated with CBGs on a piezoelectric substrate for the strain-induced tuning. All tested devices were characterized at a fixed voltage, after which the applied voltage was changed to the next fixed value to induce spectral shifts of the emission. $Al_2O_3$ layer applied by atomic layer deposition coated one half of the sample (≈2000 CBGs). For analysis we selected the CBGs with cavity mode in the spectral range from 880 nm to 920 nm and in which the trench thickness allowed coalesce of the CBG walls (i.e., complete filling of the trenches). The measurements at the closest uncoated CBGs and between CBGs provided reliable comparison of tunability. The layer configuration, devices layout, and scanning electron microscopy (SEM) images of coated and uncoated devices are presented in **Figure 1** and described in the Experimental section.

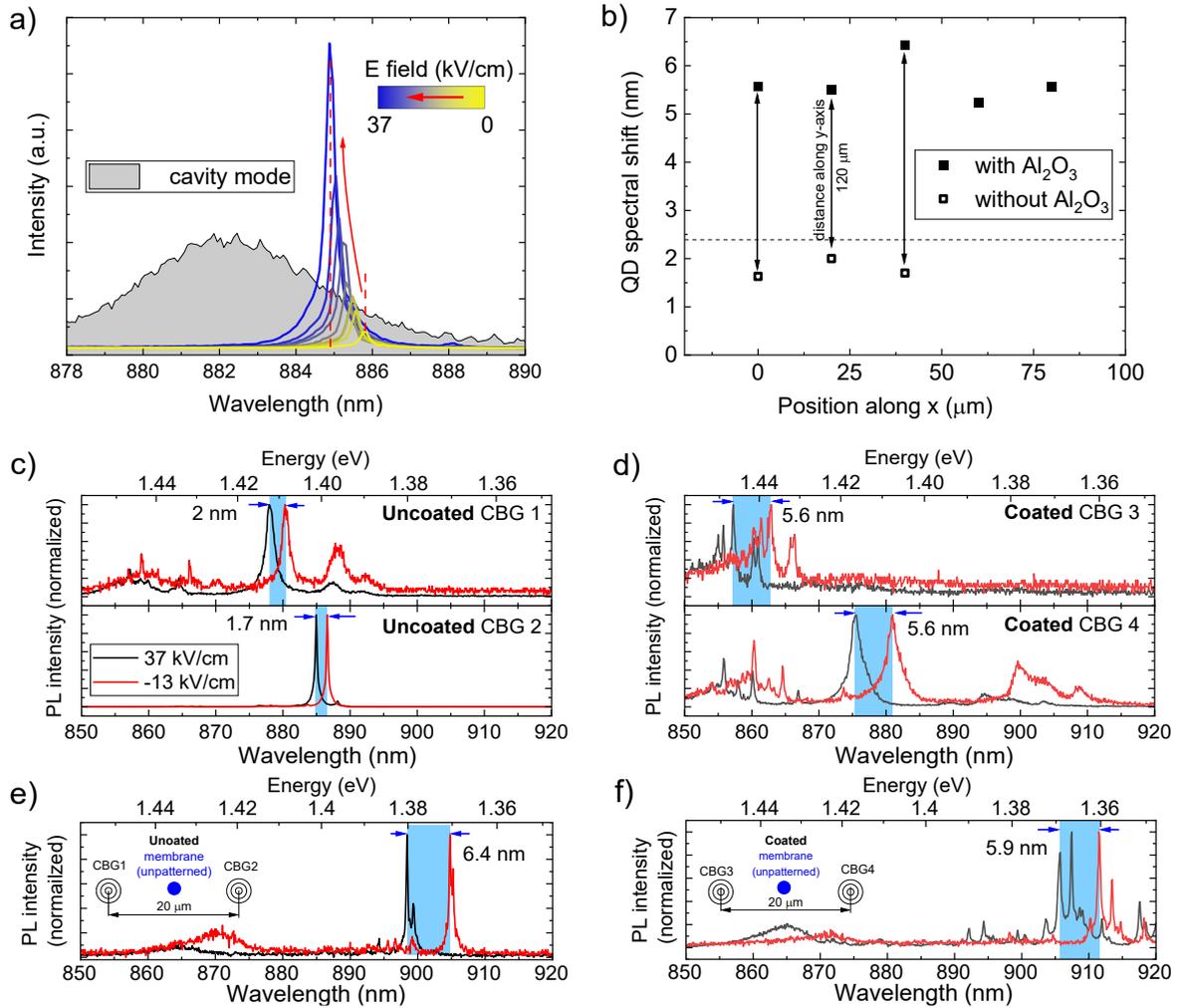

**Figure 2.** (a) Typical photoluminescence spectra for a selected QD in uncoated CBG at different voltages. The position of the cavity mode in the white light measurements is shown in grey. (b) Maximum spectral shift at full range (50 kV cm$^{-1}$) actuation is shown for 8 devices from neighboring coated (filled symbols) and uncoated fields (open symbols). CBGs in this group have trenches width 130-145 nm according to SEM ($G$ = 80). For statistical significance we measured 10 more uncoated devices in the nearest fields; all showed spectral tuning range under 2.4 nm (dashed line). Photoluminescence spectra from (c) CBG1, CBG2 and (d) CBG3, CBG4 are shown at two extreme voltages on the piezoelectric actuator. Additional spectra captured from the monolithic membrane between selected nearest devices, (e) uncoated CBG1 and CBG2 or (d) coated CBG3 and CBG4.

The spectral shift of photoluminescence was measured for quantum dots located in both coated and uncoated regions of the sample (**Figure 2**). Positive voltage on the piezoelectric actuator (aligned with the poling direction) induces a blueshift of the emission peaks. Conversely, negative voltages can result in a redshift of the emission peak. Representative emission spectra are shown in different colors in Figure 2 (a) only for positive voltage. In Figure 2 (b) we show the spectral tuning range for different devices, 5 coated and 3 uncoated CBGs. These CBGs are located in the upper two fields crossed by the dashed line in Figure 1 (d). Selected spectra from a pair of uncoated devices (CBG1, CBG2) and a pair of coated devices (CBG3, CBG4) are shown in Figure 2 (c) and Figure 2 (d), respectively. Spectral shifts related to areas of monolithic membrane between corresponding devices are shown in Figure 2 (e, f).

All devices associated with Figure 2 (b – f) were measured for a single cycle of static actuation in full voltage range from 37 kV cm$^{-1}$ to -13 kV cm$^{-1}$ to exclude the factor of piezoelectric hysteresis and variations in pre-experiment poling of the piezoelectric crystal. We checked 13 conventional devices from 3 neighboring fields in the uncoated region and found the maximum tuning range to be 2.4 nm at 1.73 nm average for the group. In the same fields, 4 points between CBGs showed average tuning range 6.48 nm at minimum result equal to 6.0 nm. Figure 2 (e) shows the spectra from unprocessed membrane with the spectral shift of 6.4 nm while the two conventional CBGs in 10 μm away this point showed the tuning range 1.7 and 2.0 nm (Figure 2 (c)). The nearest measured 5 coated CBGs ($G$ = 80 nm) demonstrate average tuning range of 5.7 nm, enhanced approximately 3 times due to complete Al$_2$O$_3$ filling of the trenches. Two of these CBGs are shown in Figure 2 (d). We notice that the unprocessed membrane keeps high spectral tuning range under the Al$_2$O$_3$ film (cf. Figure 2 (f)). Thus, the data set indicates the drop of the tunability for the conventional CBGs while the optimized cavities keep it high.

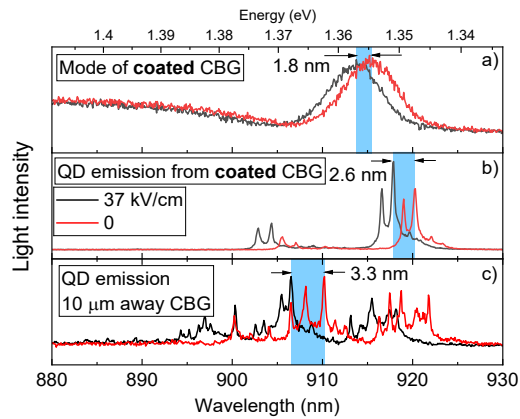

**Figure 3.** White light spectra and QD emission from a selected CBG at 0 and 37 kV cm$^{-1}$ applied voltage on the piezo electrodes. If compared to the spectral shift for QDs 10 μm away from the CBG (3.3 nm), the 1.8 nm blue shift of the CBG mode corresponds to 54% while the spectral shift of QD emission lines in this CBG is 79% (2.6 nm). For uncoated CBGs, the strain-induced shift stays always under 1/10 of corresponding QD tuning range (not shown).

In addition to the widened tuning range of the QD emission in coated CBGs, the CBG cavity modes become sensitive to the piezoelectric tuning. The spectra in **Figure 3** (a) captured for another cycle of actuation shows the cavity mode of the selected coated CBG. To specify the range of tuning, in Figure 3 (b) we demonstrate the QD emission embedded in the CBG, and (c) a random QDs' emission 10 μm away the CBG for this same actuation cycle. The monolithic membrane in this region of the sample shows the tuning range of 3.3 nm for the applied voltage range. The tuning range of the QD within the CBG is 2.6 nm and the shift of the CBG mode is 1.8 nm. For uncoated CBGs, the cavity mode tuning range is always under 1/10 of the QD's observed emission shifts. In the coated CBG, the mode tunes in the same direction as the QD, but with only 60-70% of the slope, such that the QD can still be brought in resonance at initial detuning. Relatively small initial energy difference can be achieved by deterministic placement of the CBGs, which was not performed in this study, but is in principle possible[37–39]. The simultaneous shift of both the cavity mode and the QD emission indicates that the CBG-QD system remains coupled over a wider wavelength range during the piezoelectric tuning if compared to virtually pinned mode position in conventional CBGs.

## 3. Finite element method simulations

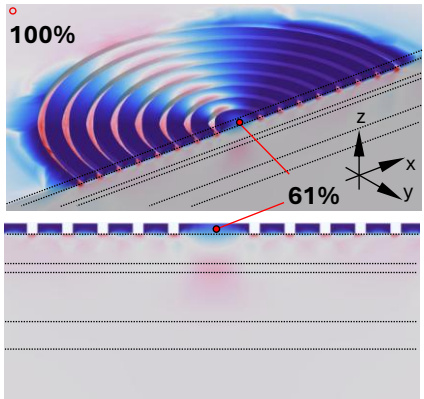
a) **Uncoated,** stiff dielectric, stiff glue

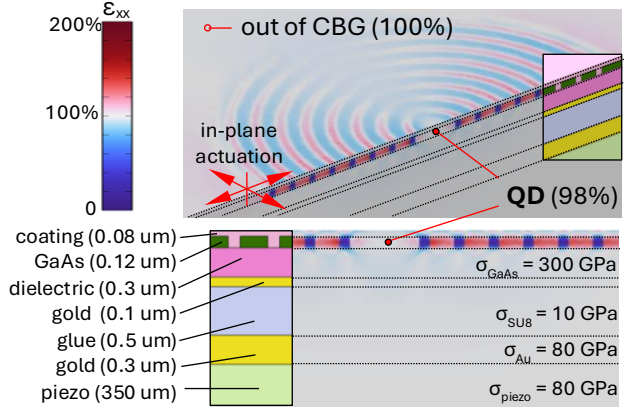
b) **Coated,** stiff dielectric, stiff glue

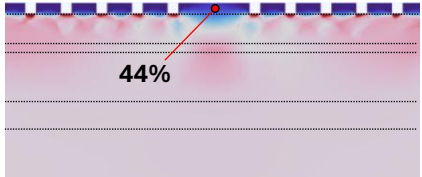
c) **Uncoated,** flexible dielectric, stiff glue

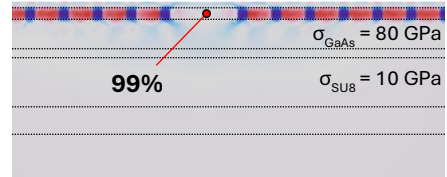
d) **Coated,** flexible dielectric, stiff glue

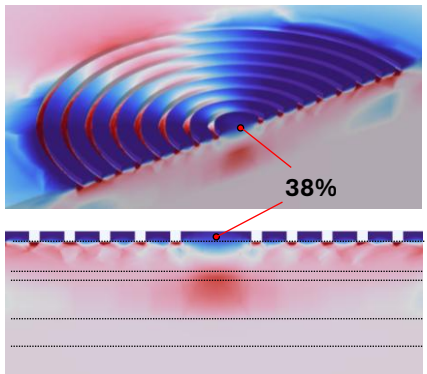
e) **Uncoated,** flexible dielectric, soft glue

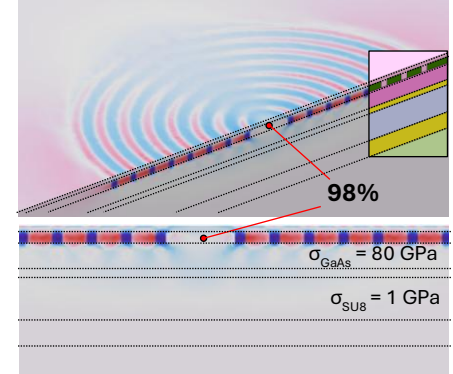
f) **Coated,** flexible dielectric, soft glue

**Figure 4.** FEM simulations of strain distribution in uncoated (a, c, e) and coated (b, d, f) CBGs. All strain values are normalized to the maximum deformation on the surface of monolithic membrane in (b) away from CBG (100% is shown in grey). The direction of the in-plane actuation is shown by arrows. The probe values (red probe point) given in % for each figure, correspond to the relative strain in the core center at $z$ in the middle of GaAs. The regions of reduced deformation (strain relaxation) appear in blue, while regions of enhanced deformation (strain accumulation) appear in red on the color maps. Compared to the configuration of materials in our measured devices (a, b), Young's modulus values between the QD and the piezoelectric substrate decreased in dielectric layer from 300 GPa to 80 GPa (c, d) imitating behavior of $SiO_2$ [34] and in the glue layer from 10 GPa to 1 GPa (e, f) imitating softer glue media.

Finite element method (FEM) simulations in COMSOL Multiphysics[40] reproduce the behavior of both uncoated and coated CBG designs (**Figure 4**). The CBGs were simulated with the following parameters: $R = 800$ nm, $P = 400$ nm, $G = 100$ nm, thickness of GaAs membrane 120 nm. The materials' Young's moduli ($\sigma$) are assumed to be close to the material properties at room temperature. [41–44] We avoid numerical evaluation of absolute material deformation by normalization of all results to the probe values on the surface of the membrane away of the CBG structure on Figure 4 (a, b). Thus, value of 100% (shown by white/grey) of the in-plane strain component $\varepsilon_{xx}$ spreads to most of the volume and major part of the top surface of the device except the CBG structure and the close area around it. The red (blue) coloration indicates strain accumulation (release) during the actuation, independent of whether the actuation corresponds to contraction or expansion of the piezoelectric substrate. The deformation $\varepsilon_{xx}$ of the bulk piezoelectric material predefined by applied voltage multiplied to piezoelectric constant $d_{31}$ of the material was found between 101 and 106% for all simulations.

Figure 4 compares coated and uncoated devices for three different configurations of $\sigma$ in the layers. The left column (Figure 4 (a, c, e)) shows uncoated devices, while the coated ones are in the right column (Figure 4(b, d, f)). The case of a stiff dielectric ($\sigma_{Al2O3} = 300$ GPa) combined with a stiff glue layer in (a, b) illustrates the strain distribution in the experimental sample studied in this paper. In Figure 4 (c, d) we reduce $\sigma$ in the dielectric coating. The change at gradual change of $\sigma$ in the coating layer but only in the probe point (red point in Figure 4 (b)) is considered in **Figure 5** (a). Since the Young's modulus of SU-8 at cryogenic temperatures is unknown to us, we assume for upper limit restriction $\sigma_{SU8} = 10$ GPa as a reasonable overestimation based on its room-temperature value of 4 GPa [42]. However, in Figure 4 (e, f) we consider additionally the glue layer at lower $\sigma = 1$ GPa. Finally, the cross-section view in Figure 5 (b) depicts the distribution of $\varepsilon_{xx}$ along the center axis of the CBG in all cases presented in previous figures.

Bright blue regions in Figure 4 (a), visible around the trenches of the CBG pattern, indicate a strong influence of the latter on strain distribution. In particular, trenches not only induce the strain relief in the rims but also initiates a macroscopic deviation of the material from the behavior of a monolithic membrane. The strain relief in the GaAs layer in CBG accompanied with the strain accumulation in the dielectric and glue layers beneath the centre of the structure lead to a bending-like deformation of the material. This mechanism is responsible for the drop of tunability of an uncoated CBG and strongly depends on the Young's modulus of the underlying layers. Depending on the device configuration and choice of materials, the strain transfer to the centre of the CBG core can be limited by 38 – 61 % if the QD layer is exactly in the middle of the etched depth. The efficiency can be under 30% (cf. Figure 5 (b)), in case if QDs are located closer to the top surface.

The effect of ALD $Al_2O_3$ coating on the strain distribution is demonstrated in Figure 4 (b, d, f). The probe values are in the range of 98-99% for all considered stiffness variations. The distribution of the strain along the CBG axis becomes homogeneous as illustrated in Figure 5 (b). It confirms that the coating suppresses parasitic bending of the layers. From a practical point of view, this indicates that the QDs in coated devices can be located at any depth under the surface for effective tuning.

In uncoated CBGs, Young's modulus in the underlayers change the resulting stress in the QD layer. According to the simulations, a stiffer underlying dielectric $Al_2O_3$ layer causes lower losses than more flexible materials such as $SiO_2$ (Figure 4 (a, c)). Both materials, are conventional choice for devices [34]. At first sight, a preference of $Al_2O_3$ in strain-induced tunable CBG devices may seem counterintuitive, since by a wrong analogy to monolithic membrane, a stiff layer is expected to oppose mechanical actuation better. We evaluate the in-plane axial stiffness of the passive composite layer on top of the piezoelectric substrate to quantify the possible impact of $Al_2O_3$ layer into the losses. For the thicknesses and moduli listed in Figure 4 (b), the stiffness of the composite film reaches 160 kN m$^{-1}$ (or 160 GPa µm, i.e., modulus by thickness), which is only 0.6% of the piezoelectric layer stiffness. Thus, in the considered devices, the 300-nm $Al_2O_3$ layer of the reflector is too thin to oppose the forces generated by the 350-µm piezoelectric substrate. The same is concluded for 80 nm top ALD $Al_2O_3$ coating. In Figure 4, the passive layers reduce the maximum strain generated by the piezoelectric substrate by no more than 6%, predominantly due to the impact of the electrode and the reflector but not the additional coating. Correction to cryogenic parameters of materials is required.

On the top x-axis in Figure 5 (a) we give the axial in-plane stiffness of the composite passive layer calculated for the reflectors at $\sigma$ = 300 GPa and $\sigma$ = 80 GPa. As is seen on the top x-axis, $\sigma$ coating over 10 GPa impacts the stiffness of composite passive layers at both flex (green) or stiff (red) reflector dielectric. Only virtual materials with an unrealistically high $\sigma$ above 1000 GPa could provide noticeable mechanical opposition. For example, $\sigma$ of $10^4$ GPa at coating film thickness of 80 nm provide 936 kN m$^{-1}$ and would cause 25% actuation losses in a monolithic membrane. This loss also limits the optimization of stress transmission through the coating. We hope this evaluation serves as a starting point for controlling coating thickness and stiffness in tunable devices.

We see good agreement of the calculated and measured efficiency of the actuation. Overall, we may recommend stiffer layers beneath the membrane and above the piezoelectric for conventional CBG devices. The $Al_2O_3$ top coating solves two problems simultaneously: it minimizes the losses on strain transfer and makes the stiffness of underlayers virtually unrelated to the efficiency of the device tuning. Noticeably, significant losses appear due to the bending-like deformation of CBGs however, the coating itself does not oppose the actuator due to neglectable thickness relative to the piezo substrate.

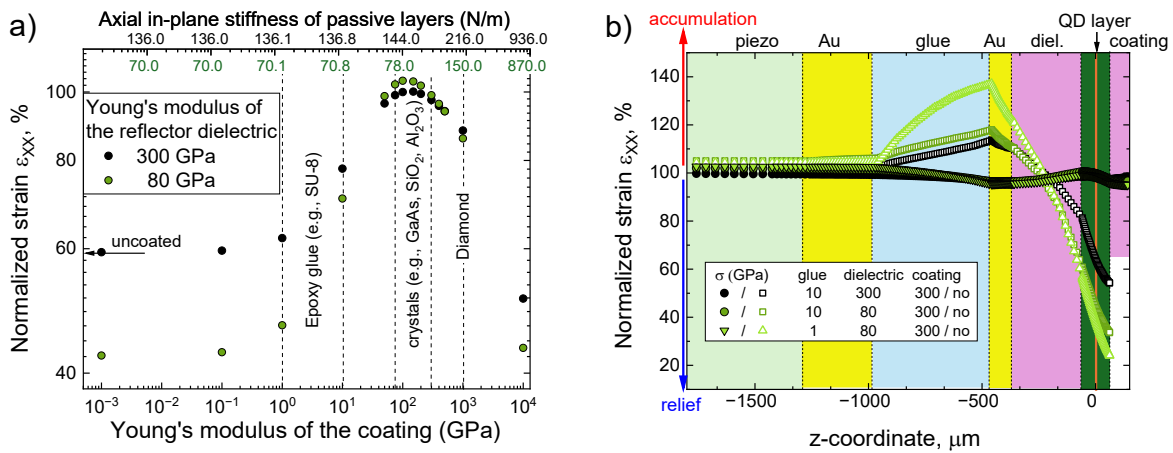

**Figure 5.** (a) The values of normalized in-plane strain ($\varepsilon_{xx}$) as a function of Young's module $\sigma$ in the coating material are shown for sample probe points from Figure 4. The variations of $\sigma$ in the layers are shown in GPa in legend. Two different reflector dielectric materials are investigated. The axial in-plane film stiffness is calculated for the composite film consisting of the passive layers on top of the piezo actuator. The configuration of the passive layers is shown in the cross-section in (b). The values of $\varepsilon_{xx}$ along the CBG symmetry axis ($z$-coordinate) are shown for all the cases depicted in Figure 4. The filled (open) symbols correspond to coated (uncoated) samples. The values above (below) 100% correspond to strain accumulation (relief).

## 4. Conclusion

In conclusion, coated CBGs represent an appropriate choice of cavities for piezoelectric strain tuning. $Al_2O_3$ coating of devices provides enhanced spectral tunability of QD emission compared to conventional CBGs. FEM simulations indicate that the strain distribution along the depth in the optimized coated CBGs closely resembles that of a monolithic membrane. Experimentally, the strain-induced spectral shifts in coated CBG devices reach 91±4% of the shifts observed in monolithic layers.

In contrast, standard CBGs without coating show a pronounced reduction in tunability. FEM predicts a drop by a factor of 2 when stiff glue and reflector underlayers are used, and by a factor of 3 for softer configurations. As shown, strain relief on the material's edges in the CBG rings and core causes parasitic bending in the device layers. This causes the leak of strain in the middle of the CBG core. The experimental data confirms a reduction by about a factor of 3. The simulations further clarify that the stiffness of glue and reflector layers strongly affects uncoated devices but has little influence in coated CBGs or monolithic membranes. Finally, the dielectric coating also modifies the behavior of the CBG cavity modes. In coated devices, the cavity mode shifts in the same direction as the QD emission, covering 60–70% of its tuning range. In uncoated CBGs, the range remains below 10%. This correlated tuning of emitter and cavity mode may be exploited in applications requiring spectral alignment across multiple devices, for instance in multi-party interference experiments in quantum networks.

## 5. Experimental

A 125 nm thick GaAs membrane (dark green in Figure 1(a)) with QDs in the middle of the layer is grown on a GaAs substrate by molecular beam epitaxy (MBE). Then a dielectric spacer layer of $Al_2O_3$ and a gold layer acting as a back reflector are applied using sputter deposition and evaporation. Next, the sample is transferred to a piezoelectric substrate via a flip-chip process, and the original GaAs wafer is removed by lapping and chemical wet etching. The piezoelectric substrate is a commercially available lead magnesium niobate-lead titanate (PMN-PT) monocrystalline piezoelectric with a nominal piezoelectric constant $d_{33}$ in the range of 1800–2000 pm V$^{-1}$ at room temperature (corresponding to $d_{31}$~800 pm V$^{-1}$). The choice of material is reasoned by high piezoelectric coefficients compared to alternatives.[45] Cr/Au layers, referred to as electrodes, with a thickness of 10/300 nm are deposited on the 350 μm PMN-PT substrate to apply voltage. Prior to the experiment, the PMN-PT was poled at room temperature[46] with a positive polarity (along with the (001) orientation of the crystal.

SU-8 (light blue in Figure 1) is employed for bonding; its thickness varies between 0.5 and 0.9 μm in different sample areas. Subsequently, we use electron beam lithography and inductively coupled plasma etching to pattern the CBGs (Figure 1(b)) in the membrane. The CBGs are fully etched through the membrane. Finaly, another $Al_2O_3$ layer (coating) 82 nm thick is applied via atomic layer deposition (ALD), covering only the right half of the sample (see Figure 1(c, d, e)). The ALD ensures that the coating fills the etched trenches of the CBGs completely as shown in Figure 1 (e). The QD layer lies in the middle of the GaAs membrane at total depth of 62.5 nm beneath the top surface (144.2 nm for the coated part).

The electron beam layout consists of 9 identical fields (Figure 1 (c)) Within a single field, different parameters are changed: 5 various radii $R$ (increase from left to right, from 380 nm to 450 nm), 3 various gap width parameter $G$ (increase from up to down, from 80 nm to 100 nm), while the period $P$ is fixed (380 nm). Each set of parameters is represented by 25 CBGs. An increase of $R$ and decrease of $G$ have provided the expected red shift of the cavity mode wavelength $\lambda$ [47,48].

The micro photoluminescence (PL) measurements are performed using a continuous wave laser at 671 nm for above-bandgap excitation and a resonant excitation of the cavity mode using a broad band (white-light) excitation source. The signal is collected using fiber in the detection path aligned near normal incidence. The spot size from which light is collected is less than 1 µm² in both experiments. The laser and the white light beams were linearly polarized along the $x$-axis, while the emitted light reaching the collecting optical fiber was always polarized along the $y$-axis (Figure 1 (b)). Using this excitation geometry the cavity mode can be resonantly excited using the white-light, which makes it easy to identify the CBG mode itself without observing QD emission. This polarization was fixed for all experiments, consequently this work does not include a polarization analysis. The laser beam was coaligned to the collection and focused on the area matching the probing area while the white light beam was slightly tilted from the $z$-axis and intentionally misfocused. An additional aperture cuts excessive white light in real and k-space allowing observation of reproducible Fano-like curves.[49–51] This way, the cavity mode from selected CBGs in the white-light experiments reveals the spectrum similar to high power pumping of QDs to fill the cavity mode, without the additional heating of the sample. An example of cavity mode as well as measured PL can be seen in Figure 2 (a) and Figure 3.

**Acknowledgements**
The authors acknowledge the support of the state of Bavaria and the Federal Ministry of Research, Technology and Space (BMFTR) within Projects QR.X (FKZ: 16KISQ010) and QR.N (FKZ: 16KIS2209). This work was supported by the Vector Stiftung (Grant No. P2024-0772), Baden-Württemberg, Germany. THL and MS acknowledge financial support from the BMFTR within the Project Qecs (FKZ: 13N16272). The authors acknowledge support from Sabrina Estevam, Monika Emmerling, Silke Kuhn, and Johannes Düreth for sample preparation.

**Data availability statement**
The data that support the findings of this study are available from the corresponding author upon reasonable request.

**Conflict of interest disclosure**
The authors have no competing financial or non-financial interests to declare.

**Enhanced spectral range of strain-induced tuning of quantum dots in circular Bragg grating cavities**

Experimental measurements and finite element simulations together reveal how stiff coating restores strain tunability in circular Bragg grating (CBG) quantum dot devices. Filling the CBG trenches with $Al_2O_3$ coating recovers up to 98% of the planar strain ($\varepsilon_{xx}$) while preserving cavity performance, enabling bright and tunable quantum light sources.

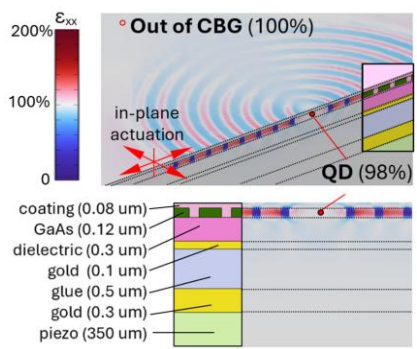